\documentclass[journal,final]{IEEEtran}
\usepackage{graphicx}
\usepackage{epstopdf}
\usepackage{setspace}
\usepackage{indentfirst}
\usepackage{enumerate}
\usepackage{color}
\usepackage{amsmath}
\usepackage{amssymb}
\usepackage{amsthm}
\usepackage{romannum}
\usepackage{blkarray}
\DeclareMathAlphabet\mathbfcal{OMS}{cmsy}{b}{n}
\newtheorem{theorem}{Theorem}
\newtheorem{remark}{Remark}

\definecolor{col}{RGB}{0,0,0}

\ifCLASSINFOpdf
\else
\fi

\begin{document}
%
\title{A Tensor-BTD-based Modulation for Massive Unsourced  Random Access\thanks{This work was supported  by the National Natural Science Foundation of China under Grant 11771244 and Grant 12171271. \textit{(Corresponding author: Liping~Zhang.)}}}%
%
%
%
\author{Zhenting~Luan\thanks{Zhenting~Luan and Liping~Zhang are with the Department of Mathematical Sciences, Tsinghua University, Beijing 100084, China (e-mail: luanzt18@mails.tsinghua.edu.cn; lipingzhang@tsinghua.edu.cn).},~Yuchi~Wu\thanks{Yuchi~Wu,~Shansuo~Liang,~Wei~Han,~and~Bo~Bai are with Theory Lab, Central Research Institute, 2012 Labs,  Huawei  Technologies Co., Ltd. (e-mail: \{wu.yuchi,  liang.shansuo, harvey.hanwei,  baibo8\}@huawei.com).},~Shansuo~Liang,~Liping~Zhang,~Wei~Han,~and~Bo~Bai}%

%

%
%

\markboth{}%
{Shell \MakeLowercase{\textit{et al.}}: Bare Demo of IEEEtran.cls for IEEE Journals}
%



\maketitle

\begin{abstract}
In this letter, we propose a novel tensor-based modulation scheme for massive unsourced random access. The proposed modulation can be deemed as a  summation of third-order tensors, of which the factors are representatives of  subspaces. A constellation design based on high-dimensional Grassmann manifold is  presented for information encoding. The uniqueness of tensor decomposition provides theoretical guarantee for active user separation. Simulation results show that our proposed method outperforms  the    state-of-the-art tensor-based modulation.
\end{abstract}


\section{Introduction}\label{intro}

\IEEEPARstart{T}{he} recent trends in the next generation cellular network and Internet-of-Things (IoT) feature increasing connectivity density of wireless devices, which motivates the development of massive random access schemes \cite{iot,tbm3}. A typical scenario is modeled as a single receiver with multiple antennas serving massive users that are sporadically active. Within a single time slot, only a small portion of the users are transmitting short-packet information payloads  simultaneously. The associated scheduling overhead renders the conventional four-step grant-based transmission inefficient and impractical. Therefore, a \textit{grant-free} communication system without pre-allocation of resources is desirable. In recent years, \textit{unsourced random access} (URA), a novel grant-free scheme where massive devices sporadically transmit signals with the same  codebook,  has drawn much attentions  \cite{tbm3}. \textcolor{black}{At the receiver, the task is to decode a list of messages transmitted by the active users. The user identifications can be embedded into the transmitted messages, thus the receiver can identify different users from the decoded message list. } Due to this feature, the performance of URA merely depends on the number of active users.


Recently, various schemes have been proposed to support massive URA. The sporadic nature of URA can be modelled as a compressed-sensing problem, where  coded compressed sensing  is proposed as the underlying coding schemes \cite{tamu}. Moreover, it has been shown that sparse regression codes achieve the symmetric multiple access channel capacity with approximate message passing decoding \cite{tbm5}. Considering quasi-static Rayleigh fading, the fundamental limits are derived for URA and a scheme based on low-density parity-check (LDPC) code is proposed to approach information-theoretic bounds under belief-propagation decoding \cite{tbm4}.

In this letter, we consider a tensor-based solution for URA. \textcolor{col}{For the single-input single-output scenario, M-PSK is factorized as the Kronecker product of lower or equal order PSK constellations \cite{R1}. The resulting constellation symbols can be treated as rank-one tensorized data blocks, where the power method are used for decoding.}  A tensor-based modulation (TBM)  with canonical polyadic decomposition (CPD) is proposed, which applies tensor lower-rank approximation for multi-user multiplexing and user separation~\cite{tbmarxiv}. Based on the high upper bound for tensor rank guaranteeing uniqueness of CPD, TBM can accommodate a large number of active users. However, the Kronecker product of vectors in TBM incurs excessive redundancy of information bearing symbols and requires large number of channel uses for a single transmission. Due to the limited time-frequency resource, such a requirement may be unfriendly to bursty transmissions in  high-speed mobile environments. To cope with this, we propose a novel modulation design based on block term decomposition (BTD) of tensors \cite{btdpart2} for URA. At the transmit side, each user sends a signal sequence of the structured tensor, of which the factor matrices are designed with high-dimensional Grassmann manifold. At the receiver, a base station (BS) equipped with multiple antennas employs BTD to retrieve all users' information bits non-coherently.   The main contributions are summarized as follows.
\begin{itemize}
    \item \textcolor{black}{We propose a novel encoding and decoding scheme based on Grassmann manifold, which regards matrix symbols as representatives of high-dimensional subspaces.}
    \item Spectrum  efficiency of the proposed modulation scheme is higher than TBM since  the encoded symbols with high-dimensional subspaces can carry more information bits than the same-length symbols in TBM.
    \item The user-grouping strategy is beneficial  for decreasing the demodulation complexity, reducing the error rate of signal recovery, as well as enabling more flexible resource scheduling due to a more  refined channel resource granularity.
\end{itemize}


\section{Backgrounds on URA Model and Tensor-Based Modulation}\label{sectionmura}

\subsection{System Model}\label{sectionsystemmodel}

We consider an uplink multi-user grant-free system, where $M$ single-antenna users communicate with an $N$-antenna BS. We assume that there are only $K\!\ll\!M$ active users within a time block of $T_c$ channel uses.  Denote $\textbf{s}_k\in \mathbb{C}^{T_c\times 1}$ and $\textbf{h}_k\in \mathbb{C}^{N\times 1}$  as  the $k$-th user's transmitted signal and its channel gain, respectively. Then, the received signal $\textbf{Y}\in \mathbb{C}^{T_c\times N}$ at BS is given by
\begin{equation}\label{receivesignalgeneral}
    \textbf{Y}=\sum_{k=1}^K\textbf{s}_k\textbf{h}_k^\mathsf{T}+\textbf{N},
\end{equation}
where $\textbf{h}_k$ is assumed to contain independent and identically distributed (i.i.d.) complex Gaussian entries with unit variance and $\textbf{N}\in \mathbb{C}^{T_c\times N}$ is the additive Gaussian white noise (AWGN).   At the receive side, the non-coherent detection is developed to recover  $\textbf{s}_k$ with unknown $\textbf{h}_k$.

\subsection{Review of Tensor-Based Modulation for URA Model}

In this subsection, we briefly review a tensor-based modulation (TBM) scheme for the URA problem  \cite{tbmarxiv}. Specifically, the transmitted signal $\textbf{s}_k$ of $k$-th user was given by  a rank-$1$ tensor of dimensions $T_1,T_2,\ldots,T_d$:
\begin{equation}\label{tbmsignal}
    \textcolor{black}{\textbf{s}_k}=\textbf{x}_{1,k}\otimes\textbf{x}_{2,k}\cdots\otimes \textbf{x}_{d,k},
\end{equation}
where $\otimes$ denoted the tensor product and $T_c=\prod_{i=1}^d T_i$. Here,  the $i$-th factor  $\textbf{x}_{i,k}\in \mathbb{C}^{T_i\times 1}$ was drawn from a sub-constellation~$\mathcal{C}_i$ shared by all users. {\color{black} For narrative convenience, we represent faded signals in tensor form with capital italic bold letters in the rest of this letter.  Then, the received signal $\textbf{\textit{Y}}$ in (\ref{receivesignalgeneral}) can be rewritten in a $(d\!+\!1)$\textsuperscript{th}-order tensor form, i.e.,
\begin{equation}\label{ykform}
    \textit{\textbf{Y}}=\sum_{k=1}^K\textbf{s}_k\otimes\textbf{h}_k+\textit{\textbf{N}}
=\sum_{k=1}^K(\textbf{x}_{1,k}\otimes\cdots\otimes \textbf{x}_{d,k})\otimes\textbf{h}_k +\textit{\textbf{{N}}},
\end{equation}
where $\textit{\textbf{Y}}$ was a summation of $K$ rank-$1$ tensors in the CPD structure corrupted by noise $\textit{\textbf{N}}$.}

At the receiver, a two-step detection method was proposed to recover all users' signals $\textbf{s}_k$ \cite{tbmarxiv}. First, a tensor CPD algorithm was employed to  retrieve all rank-$1$ tensors, \textcolor{black}{where the uniqueness of CPD is addressed in \cite[Theorem 1.1]{cpduniqueness}}. Second, the factor vectors of the recovered rank-$1$ tensors were demapped to information bits. To ensure the accuracy of demapping, the Grassmann constellation $\mathcal{C}_i$ consisted of all one-dimensional subspaces in $\mathbb{C}^{T_i}$. This design of constellation substantially embeds a reference element into each symbol to identify the belonging subspace and correct  offset.

The TBM scheme together with the two-step detection method enables massive connectivity over the shared uplink wireless channel. However, the construction of rank-1 tensors implies an excessive reuse of factor vectors, which leads to low spectrum efficiency. In  next section, we generalize TBM to a novel modulation with more flexible tensor structures, which enables higher transmission efficiency.

\section{Modulation Design Based on  Block Term Decomposition}\label{sectionmodulation}

In this section, we propose a tensor modulation scheme based on block term decomposition (BTD) for massive URA system. We first state the main ideas of the proposed modulation, and then  introduce a specific constellation design scheme for the BTD-based modulation (BTDM).

\subsection{BTDM Scheme}\label{btdm}
\textcolor{black}{In BTDM, we utilize $L$-column unitary matrices ($L>1$),   as symbols to construct signals. The symbols are representatives of $L$-dimensional subspaces, which are elements of Grassmann manifold \cite{quotient}. Specifically, the signal of $k$-th user is transmitted in the form of a matrix, which is the product of two
full-column-rank matrices  $\textbf{A}_k\!\in\!\mathbb{C}^{T_1\times L}$ and $\textbf{B}_k\!\in\!\mathbb{C}^{T_2\times L}~ (L\!<\!T_1,T_2)$, i.e.,}
\begin{equation}\label{singlesymbol}
\textcolor{black}{\textbf{s}_k=\textbf{A}_k\textbf{B}_k^\mathsf{T}\in \mathbb{C}^{T_1\times T_2},}
\end{equation}
\textcolor{black}{where $T_c\!=\!T_1T_2$ is the number of channel uses.
Here we use $\textbf{s}_k$ to represent matrix signals, with abuse of the notation in (\ref{receivesignalgeneral}). $\textbf{A}_k$ and $\textbf{B}_k$ are \textit{symbols}   drawn from sub-constellations $\mathcal{C}_1\!\subseteq\!\mathbb{C}^{T_1\!\times\! L}$ and $\mathcal{C}_2\!\subseteq\!\mathbb{C}^{T_2\!\times\! L}$, respectively.  $\mathcal{C}_1$ and $\mathcal{C}_2$ are designed with the same method introduced in the next subsection.}


Following a similar procedure as  (\ref{ykform}), the received signal can be rewritten as a  faded signal with third-order tensor form:
\begin{equation}\label{receivedsignal} \textit{\textbf{Y}}=\sum_{k=1}^K(\textbf{A}_k\textbf{B}_k^\mathsf{T})\otimes \textbf{h}_k+\textit{\textbf{N}},
\end{equation}
where  $\textit{\textbf{Y}}_k\triangleq(\textbf{A}_k\textbf{B}_k^\mathsf{T})\otimes \textbf{h}_k$ is called a \textit{block term}  defined as a faded signal tensor transmitted by $k$-th user. The right-hand side of (\ref{receivedsignal}) follows the structure of BTD for $\textit{\textbf{Y}}$ \cite{btdpart2}. Compared to TBM,  each user's signal tensor in BTDM can accommodate more elements, which carries more information bits than TBM given the same number of symbols. This improves the spectrum efficiency.

\subsection{Constellation Design}\label{constellationdesignsubsection}

\begin{figure}[ht]
    \centerline{\includegraphics[width=9cm,height=3.2cm]{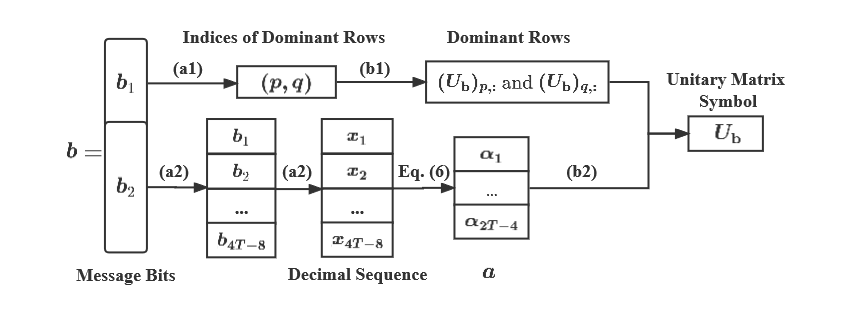}}
    \caption{Encoding message bits $\textbf{b}$ into constellation. }\label{encodingflow}
\end{figure}

In this subsection, we present a two-step constellation design for $\mathcal{C}_1$ and $\mathcal{C}_2$, and describe the constellation mapping procedures with a concrete example of $L\!=\!2$ in Fig. \ref{encodingflow}.\footnote{ Since design scheme for $\mathcal{C}_1$ and $\mathcal{C}_2$ are the same, we omit the subscript in this subsection for simplicity. Specifically, we use $T$ to represent $T_1$ and $T_2$.}  Extension to larger $L$ is straightforward.

\textcolor{col}{Given an information sequence  $\textbf{b}$ of length-$\ell$, the following procedures create a corresponding  constellation symbol $U_{\textbf{b}}\in \mathbb{C}^{T\times L}$. We split $\textbf{b}$ into two parts $(\textbf{b}_1,\textbf{b}_2)$ and encode $\textbf{b}_1$ and $\textbf{b}_2$ separately.   Let $\lfloor\cdot\rfloor$ denote the round-down operator. }
\begin{itemize}
    \item[(a1)] The first $\ell_{1}=\lfloor \log_2(\frac{T(T\!-\!1)}{2})\rfloor$ bits, denoted as $\textbf{b}_1$, will be employed to determine the \textit{dominant rows} in $U_\textbf{b}$, of which the Frobenius norms are the largest among all $T$ rows. We map $\textbf{b}_1$ into a set of integer pairs $\{(p,q)\!:\!1\!\leq\!p\!<\!q\!\leq\!T\}$ via an injection, where the latter has $\frac{T(T\!-\!1)}{2}$ pairs .
    \item[(a2)]  The  rest $\ell_{2}=\ell-\ell_1$ bits, denoted as $\textbf{b}_2$, corresponds to the rest rows of $U_\textbf{b}$. We uniformly divide $\textbf{b}_2$ into $(4T\!-\!8)$ parts once again, i.e., $\textbf{b}_{2}=(\widetilde{\textbf{b}}_1,\cdots,\widetilde{\textbf{b}}_{4T-8})$, then deem each part $\widetilde{\textbf{b}}_i$ as a binary number and transform it into decimal number $x_i(i\!\leq\!4T-8)$.  Via \cite[(12)]{cubesplit}, the decimal sequence $(x_1,\cdots,x_{4T-8})$ is mapped into a complex vector  $\textbf{a}=(\alpha_1,\cdots,\alpha_{2T-4})^\mathsf{T}$ with half length, of which the elements are uniformly distributed inside a unit complex circle. Specifically,  the mapping is
\begin{equation}\label{phi}
    \alpha_t=\sqrt{\frac{1-e^{-\frac{|\omega_t|^2}{2}}}{1+e^{-\frac{|\omega_t|^2}{2}}}}\frac{\omega_t}{|\omega_t|},\ t\in [2T\!-\!4],
\end{equation}
where $\omega_t=\mathcal{N}^{-1}(\frac{2x_{2t-1}+1}{2^{L_{2t-1}+1}})+j\mathcal{N}^{-1}(\frac{2x_{2t}+1}{2^{L_{2t}+1}})$ with  $\mathcal{N}(\cdot)$ being the cumulative distribution function of standard normal distribution and $L_t$ is the length of $\widetilde{\textbf{b}}_{t}$.
\end{itemize}
\textcolor{col}{With these preparations, we build  symbol $U_\textbf{b}$ by three steps.}
\begin{itemize}
    \item[(b1)] Construct dominant rows. Denote $i_p=\textbf{a}_{1:T-2}^H\textbf{a}_{T-1:2T-4}$ as the inner product between two halves of $\textbf{a}$ and $f>\sqrt{2}$ as a  constant. Then, the dominant rows of $U_\textbf{b}$  corresponding to $\textbf{b}_1$ are designed as
    \begin{equation}\label{special}
    (U_\textbf{b})_{p,:}=(f,-i_p/f), \qquad (U_\textbf{b})_{q,:}=(0,f).
    \end{equation}
    \item[(b2)] Fill the rest entries of $U_\textbf{b}$. The rest $2T-4$ entries of $U_\textbf{b}$ are orderly filled with elements of $\textbf{a}$ from top to bottom and from left to right. Hence, the Frobenius norms of these $T-2$ rows are less than $\sqrt{2}$, thus are  less than the norms of dominant rows, as well.
    \item[(b3)] Scale $U_\textbf{b}$ to make it unitary.
\end{itemize}
\textcolor{black}{Then, $U_\textbf{b}$ will be utilized to transmit messages $\textbf{b}$ as one symbols in (\ref{singlesymbol}). In BTDM, user $k$ simultaneously transmits two symbols $\textbf{A}_k$ and $B_k$ with the composite signal $\textbf{s}_k$. }
\begin{remark}
  \textcolor{col}{  The design of dominant rows as (\ref{special}) ensures the unitariness of $U_\textbf{b}$ with proper scaling. Meanwhile, the parameter $f$ guarantees that these rows are dominant according to Frobenius norm, which is invariant with  rotation or scaling of $U_\textbf{b}$. Hence, the dominant rows  not only work as pilots, but also carry the information bits in $\textbf{b}_1$. A perspective with Grassmann manifold to this design will be illustrated in Section \ref{sectionmath}.}
\end{remark}





\section{Grassmann Manifold and Demodulation for BTDM }\label{sectionmath}

The task at the receiver is to precisely recover $K$ sub-tensors in (\ref{receivedsignal}), which requires a unique tensor decomposition of \textit{\textbf{Y}}. In this section, we first discuss the uniqueness of BTD. Then, we introduce the Grassmann manifold for constellation design and present demodulation procedures for BTDM.

\subsection{Essential Uniqueness of BTD}\label{sectionuniqueness}

In general, tensor  decomposition in  the form of (\ref{receivedsignal}) is not unique. For example, the equality  (\ref{receivedsignal}) also holds as
\begin{equation}\label{inviriance}
\textit{\textbf{Y}}_k=(\textbf{A}_k\textbf{B}_k^\mathsf{T})\otimes \textbf{h}_k=((\alpha_1\textbf{A}_k\textbf{S}_k)(\alpha_2\textbf{B}_k(\textbf{S}_k^\mathsf{T})^{-1})^\mathsf{T})\otimes(\alpha_3\textbf{h}_k),
\end{equation}
for any complex scaling factors $\alpha_1,\alpha_2,\alpha_3$ with  $\alpha_1\alpha_2\alpha_3=1$  and any nonsingular matrix $\textbf{S}_k\in \mathbb{C}^{L\times L}$.  We call the BTD essentially unique if the decomposition (\ref{receivedsignal}) is unique in the sense of (\ref{inviriance})  \cite{btdpart2}. {\color{black} The theorem below states sufficient conditions for the essential uniqueness and  provides an  upper bound of $K$, i.e., the number of active users. Details of its proof can be found in \cite[Theorem 4.1 and 4.7]{btdpart2}.}
\begin{theorem}[Essential Uniqueness \cite{btdpart2}] \label{uniqueness}
The BTD (\ref{receivedsignal}) is essentially unique when satisfies one of the following conditions.
\begin{itemize}
    \item $N\geq K~and~\min(\lfloor \frac{T_1}{L}\rfloor,K)+\min(\lfloor \frac{T_2}{L}\rfloor,K)\geq K+2;$
    \item $\lfloor\frac{T_1T_2}{L^2}\rfloor\geq K~ and ~ \min(\lfloor \frac{T_1}{L}\rfloor,K)+\min(\lfloor \frac{T_2}{L}\rfloor,K)+\min(N,K)\geq 2K+2.$
\end{itemize}
\end{theorem}

According to (\ref{inviriance}), the rotation- and scale-invariant properties is required for  constellation design. To address this issue, we  introduce Grassmann manifold in the following subsection. Moreover, based on Theorem \ref{uniqueness},  we establish degree of freedom of BTDM in the extended version of this letter \cite[Section \Romannum{4}-D]{arxivbtdm}.

\subsection{Grassmann Manifold for Constellation Design}\label{subsectiongrassmann}

The constellation design introduced in Section \ref{constellationdesignsubsection} facilitates BTD in correctly recovering the correct sub-constellation symbols and the corresponding channel fadings with the invariance property of Grassmann manifold \cite{quotient}.

The Grassmann manifold $G(T,L)$ with $T\!>\!L$ is a quotient space of $U(T,L)=\{\textbf{M}\in\mathbb{C}^{T\times L}:\textbf{M}^H\textbf{M}=\textbf{E}_L\}$, i.e., $G(T,L)=U(T,L)/\!\sim,$ where $\sim$ denotes an  equivalence relation defined as: for any $\textbf{M},\textbf{N}\in U(T,L), \textbf{M}\sim \textbf{N}$ if and only if there exists a unitary matrix $\textbf{S}\in \mathbb{C}^{L\times L}$ such that $\textbf{M}=\textbf{NS}$. Hence, $G(T,L)$ consists of all $L$-dimensional  subspaces of $\mathbb{C}^{T\times 1}$.
For each $L$-dimensional subspace, we can find a unitary matrix $\textbf{M}_0\in \mathbb{C}^{T\times L}$ with the same column space to represent it. Then, the subspace is denoted as $\overline{\textbf{M}}_0$  and $\textbf{M}_0$ is called a representative of $\overline{\textbf{M}}_0$. Moreover, different representatives of same subspace are inter-convertible with some unitary rotational matrix. Here we employ  Chordal distance introduced in \cite{distance}  for Grassmann manifold :
\begin{equation}\label{distance}
    d(\overline{\textbf{M}},\overline{\textbf{N}})=\sqrt{2-|tr(\textbf{M}^H\textbf{NN}^H\textbf{M})|}, ~\forall~ \overline{\textbf{M}},\overline{\textbf{N}}\in G(T,2),
\end{equation}
where $\textbf{M},\textbf{N}$ are  representatives of $\overline{\textbf{M}},\overline{\textbf{N}}$, respectively.

Similarly to Section \ref{constellationdesignsubsection}, we illustrate the relation between Grassmann manifold and the proposed constellation design in the case of $L=2$. \footnote{ We  omit the subscript of $T_1$ and $T_2$ and uniformly replace with $T$ like Section \ref{constellationdesignsubsection}.} Obviously, the proposed three-step design of matrix symbols in Section \ref{constellationdesignsubsection} ensures that each symbol is a representative  of the element in $G(T,2)$.

To interpret the function of the dominant rows (\ref{special}) in our proposed constellation design, we first partition $G(T,2)$ into $T(T\!-\!1)/2$ regions with a set of Grassmann points $\{\overline{\textbf{G}}^{(p,q)}\}_{1\leq p<q\leq T}$. Specifically, the  representative elements $\textbf{G}^{(p,q)}\in \mathbb{C}^{T\times 2}$ are defined as
\begin{equation}\label{cellpoint}
\textbf{G}^{(p,q)}_{ij}=\left\{\begin{aligned}
    1,& (i,j)\!=\!(p,1) ~\text{or}~(q,2) \\
    0,& ~\text{otherwise}
\end{aligned}, \forall i\in [T],  j\in [2],\right.
\end{equation}
where $[T]=\{1,2,\cdots,T\}.$ The region $\mathcal{R}_{p,q} (1\leq p<q\leq T)$ is defined as the set of all subspaces closest to $\overline{\textbf{G}}^{(p,q)}$, i.e.,
\begin{equation}\label{region}
    \mathcal{R}_{p,q}=\{\overline{\textbf{M}}\in G(T,2) : d(\overline{\textbf{M}},\overline{\textbf{G}}^{(p,q)})=\min_{1\leq s<t\leq T} d(\overline{\textbf{M}},\overline{\textbf{G}}^{(s,t)})\}.
\end{equation}
Hence, $\{\mathcal{R}_{p,q}\}_{1\leq p<q\leq T}$ is a partition of $G(T,2)$ omitting any zero-measure intersections. We simplify the Chordal distance to be minimized in (\ref{region}) into:
\begin{equation}\label{disregion}
\begin{aligned}
    d(\overline{\textbf{M}},\overline{\textbf{G}}^{(s,t)})&=\sqrt{2-|tr((\textbf{G}^{(s,t)H}\textbf{M})^H\textbf{G}^{(s,t)H}\textbf{M})|}\\
    &=\sqrt{2-\lVert \textbf{M}_{s,:}\rVert_F^2-\lVert \textbf{M}_{t,:}\rVert_F^2},
    \end{aligned}
\end{equation}
where $\lVert\cdot\rVert_F$ denotes the Frobenius norm. Then, $\mathcal{R}_{p,q}$ is the set of all $\overline{\textbf{M}}\in G(T,2)$ that the Frobenius norms of  $p$-th and $q$-th rows of its representative $\textbf{M}$ are the largest among all rows, i.e., the index pair of dominant rows is $(p,q)$. Note that this equivalence is well-defined for any representative of $\overline{\textbf{M}}$ since rotation holds the Frobenius-norm dominance of these rows.

\textcolor{black}{
According to the above discussion, the symbol building procedures introduced in Section \ref{constellationdesignsubsection} map information bits $\textbf{b}$ into a unitary matrix symbol $U_\textbf{b}$ that represents a subspace in $G(T,2)$. All such symbols constitute a \textit{Grassmann constellation}. The first step transforms  $\textbf{b}_1$ into dominant rows of $U_\textbf{b}$, which substantially allocates the subspace represented by the raw symbol into $\mathcal{R}_{p,q}\!\subseteq\! G(T,2)$. The second step completes the rest rows of $U_\textbf{b}$ by referring to $\textbf{b}_2$.
 Owing to the dominant rows, the receiver can identify the corresponding Grassmann region without detecting  symbol rotation and scaling,  which facilitates rebuilding the symbol structure as (\ref{structure}).}


\subsection{Demodulation for BTDM}\label{subsectiondemodulation}


The demodulation of received signal in (\ref{receivedsignal}) can be formulated as the following optimization problem:
\begin{equation}\label{optimization}
\{\hat{\textbf{A}}_k,\hat{\textbf{B}}_k,\hat{\textbf{h}}_k\}_{k=1}^K=\mathop{\arg\min}_{\substack{\textbf{A}_k\in \mathcal{C}_1, \textbf{B}_k\in \mathcal{C}_2, \\ \textbf{h}_k\in \mathbb{C}^{N\times 1}, \forall k\in [K]}}\lVert \textit{\textbf{Y}}-\sum_{k=1}^K(\textbf{A}_k\textbf{B}_k^\mathsf{T})\otimes \textbf{h}_k \rVert_F^2,
\end{equation}
where sub-constellations $\mathcal{C}_1$ and $\mathcal{C}_2$ are designed in Section \ref{constellationdesignsubsection}.
Solving (\ref{optimization}) can be computationally prohibitive. To tackle with the complexity, we propose a sub-optimal two-step solution. First, we decompose the received tensor signal $\textit{\textbf{Y}}$ with BTD ignoring the underlying constraint of sub-constellations $\mathcal{C}_1$ and $\mathcal{C}_2$.  We directly employ the Gauss-Newton dogleg trust-region (GNDL) algorithm  for BTD \cite{gndl} to obtain the orthonormal estimates $\hat{\textbf{A}}_k$ and $\hat{\textbf{B}}_k$. Second, the estimates  are demapped into information bits. Here we take $\hat{\textbf{A}}_k$ as an example to show the detailed demapping procedure.\footnote{ The demapping procedures for $\hat{\textbf{A}}_k$ and $\hat{\textbf{B}}_k$ are similar. Here we introduce the procedures in the case of $L=2$ and omit the subscript of $\textbf{A}_k$.}

With the essential uniqueness introduced in Section \ref{sectionuniqueness}, the estimate $\hat{\textbf{A}}$ spans the same column space as $\textbf{A}$. Hence, there exists an orthonormal matrix $\textbf{P}\!\in\!\mathbb{C}^{2\times 2}$ such that $\hat{\textbf{A}}\!=\!\textbf{A}\textbf{P}$. We sort the $2$-norm of each row of $\hat{\textbf{A}}$ and find the two largest  rows $(\hat{p},\hat{q}) (\hat{p}<\hat{q})$, which exactly correspond to the information expressed by the first $b_1$ bits in  user's message according to the analysis in Section \ref{subsectiongrassmann}.  For instance, when $(\hat{p},\hat{q})\!=\!(1,2)$,  the  original constellation symbol is expressed as
\begin{equation}\label{structure}
\textbf{A}=\begin{gathered}
\begin{pmatrix} f & -i_p/f \\ 0 & f \\ \textbf{a}_{1:T_1-2} & \textbf{a}_{T_1-1:2T_1-4} \end{pmatrix}\begin{pmatrix}
v_1^{-1} & 0 \\0 & v_2^{-1}
\end{pmatrix}\triangleq\begin{pmatrix}
\textbf{A}_{1} \\ \textbf{A}_{2}
\end{pmatrix}\textbf{V},
\end{gathered}
\end{equation}
where $f>\sqrt{2}$, $i_p=\textbf{a}_{1:T_1-2}^H\textbf{a}_{T_1-1:2T_1-4}$, $v_1$ and $v_2$ are Frobenius norms of the first and second columns of  $\textbf{A}\textbf{V}^{-1}$, respectively. Further, we denote $\textbf{A}_{1}, \textbf{A}_{2}$ as the first two rows and last $(T_1\!-\!2)$ rows of $\textbf{A}$, respectively. Then, we have
\begin{equation}\label{A1}
\begin{gathered}
\begin{pmatrix}
\hat{\textbf{A}}_{1} \\ \hat{\textbf{A}}_{2}
\end{pmatrix}
\end{gathered}=\hat{\textbf{A}}=\textbf{A}\textbf{P}=\begin{gathered}
\begin{pmatrix}
\textbf{A}_{1}\textbf{V}\textbf{P} \\ \textbf{A}_{2}\textbf{V}\textbf{P}
\end{pmatrix}
\end{gathered},
\end{equation}
hence
\begin{equation}\label{A2}
\begin{aligned}
    \textbf{A}_{2}&=\hat{\textbf{A}}_{2}(\textbf{V}\textbf{P})^{-1}=\hat{\textbf{A}}_{2}(\textbf{A}_{1}^{-1}\hat{\textbf{A}}_{1})^{-1} =\hat{\textbf{A}}_{2}\hat{\textbf{A}}_{1}^{-1}\textbf{A}_{1}\\
    &\triangleq \hat{\textbf{W}}\begin{gathered}
    \begin{pmatrix}
    f & -i_p/f \\0 & f
    \end{pmatrix}
    \end{gathered},
    \end{aligned}
\end{equation}
where $\hat{\textbf{W}}=\hat{\textbf{A}}_{2}\hat{\textbf{A}}_{1}^{-1}$ is obtained from GNDL. Therefore, we get the following one-variate linear equation for $i_p$:
\begin{equation}\label{equation}
    i_p=\textbf{a}_{1:T_1-2}^H\textbf{a}_{T_1-1:2T_1-4}=(f~0)\hat{\textbf{W}}^H\hat{\textbf{W}}\begin{gathered}
    \begin{pmatrix}
    -i_p/f \\ f
    \end{pmatrix}
    \end{gathered}.
\end{equation}
It is straightforward to obtain the solution $\hat{i}_p$ and then $\textbf{A}_{2}$ can be recovered by (\ref{A2}). Henceforth, all elements of $\textbf{a}$ are obtained and one can utilize the inverse mapping  of (\ref{phi})  to retrieve the information bits $\hat{\textbf{b}}_2$ \cite[(14)]{cubesplit}. Combining the first half bits $\hat{\textbf{b}}_1$ corresponding to the two largest-norm rows and the second half bits $\hat{\textbf{b}}_2$ obtained by (\ref{A1})$-$(\ref{equation}), we can decode complete information bits $\hat{\textbf{b}}=(\hat{\textbf{b}}_1,\hat{\textbf{b}}_2)$ from $\hat{\textbf{A}}$.

\begin{remark}
\textcolor{black}{ The Grassmann region index $(p,q)$ satisfies $1\leq p<q\leq T$. Hence, there are $\frac{T(T-1)}{2}$  Grassmann regions, which can be identified with $\log_2\frac{T(T-1)}{2}$ bits, i.e., the length of former-half bits $\textbf{b}_1$. Therefore,  different Grassmann regions correspond to different $\textbf{b}_1$ and different dominant rows,  which results in different symbol structures.}
\end{remark}

\subsection{Degrees of Freedom}
{\color{black}
In this subsection, we establish the degrees of freedom (DOF) of BTDM. Assumption of independent and identical distribution (i.i.d.) applies to both the Grassmann symbols and the channel fading.

Each user in BTDM   simultaneously transmits two symbols in $G(T_i,L),i\in [2]$. A symbol in $G(T,L)$ has $(T\!-\!L)L$ degrees of freedom (DOF) due to the unitary constraint. Hence, DOF of each user is $(T_1\!+\!T_2\!-\!2L)L$. And the total DOF in BTDM is the summation of per-user DOF:
\begin{equation}\label{dof}
    \text{DOF}_{\text{total}}=K(T_1+T_2-2L)L\leq\overline{K}(T_1+T_2-2L)L,
\end{equation}
where $\overline{K}$ is the upper bound of number of active users obtained from Theorem \ref{uniqueness}.
}

\section{Numerical Results}\label{sectionnumerical}
\subsection{Performance of BTDM and Successive  Cancellation}\label{41}

We assume that each user has a payload of $B_{0} = 204$ bits that are  encoded by a  Bose–Chaudhuri–Hocquenghem  (BCH) code. Here, the BCH code is used  for error detection and error correction. The BCH-encoded code with $B_{\text{BCH}}\!=\!220$  bits are then divided in into two parts: $B_1\!=\!124$ and $B_2\!=\!96$, which are mapped into two Grassmann sub-constellations $\mathcal{C}_1$ and $\mathcal{C}_2$, respectively. To maximize the minimum distance between  $\mathcal{C}_1$ and $\mathcal{C}_2$, we choose $T_i$ to be proportional to $B_i$ as $(T_1,T_2)=(30,24)$  \cite[Lemma 1]{cubesplit} and the  total number of channel uses is $T_c\!=T_1T_2\!=720$.
During simulations, the additive noise process is generated based on the per-bit signal-to-noise ratio $E_b/N_0$, where $E_b\!=\!\frac{\lVert\textbf{s}_k\rVert^2}{B_0}$ denotes the average energy per bit for all $k$ and $N_0$ is the power of the the AWGN. The average per user probability of error (PUPE) is defined as \cite{tbmarxiv}:
\begin{equation}\label{pupe}
    \text{PUPE}=E\left(\frac{|\mathcal{L}-\hat{\mathcal{L}}|}{|\mathcal{L}|}\right),
\end{equation}
where $\mathcal{L}$ denotes the set of messages transmitted by $K$ users and $\hat{\mathcal{L}}$ denotes the set of all messages   detected by the receiver.  In  simulations, we assume that the number of active users is known. \textcolor{black}{ In practice, the receiver processes no prior knowledge of $K$. However, it is proper to assume that $K$ is the upper bound from Theorem \ref{uniqueness}  \cite{tbmarxiv}. Then,  the receiver employs BTD to received signal tensor and discards the incorrect block terms, of which the signal power are below certain thresholds or some errors are detected by the BCH code. The remaining block terms are deemed as the active users.}

\begin{figure}[ht]
    \centerline{\includegraphics[width=9cm,height=4.8cm]{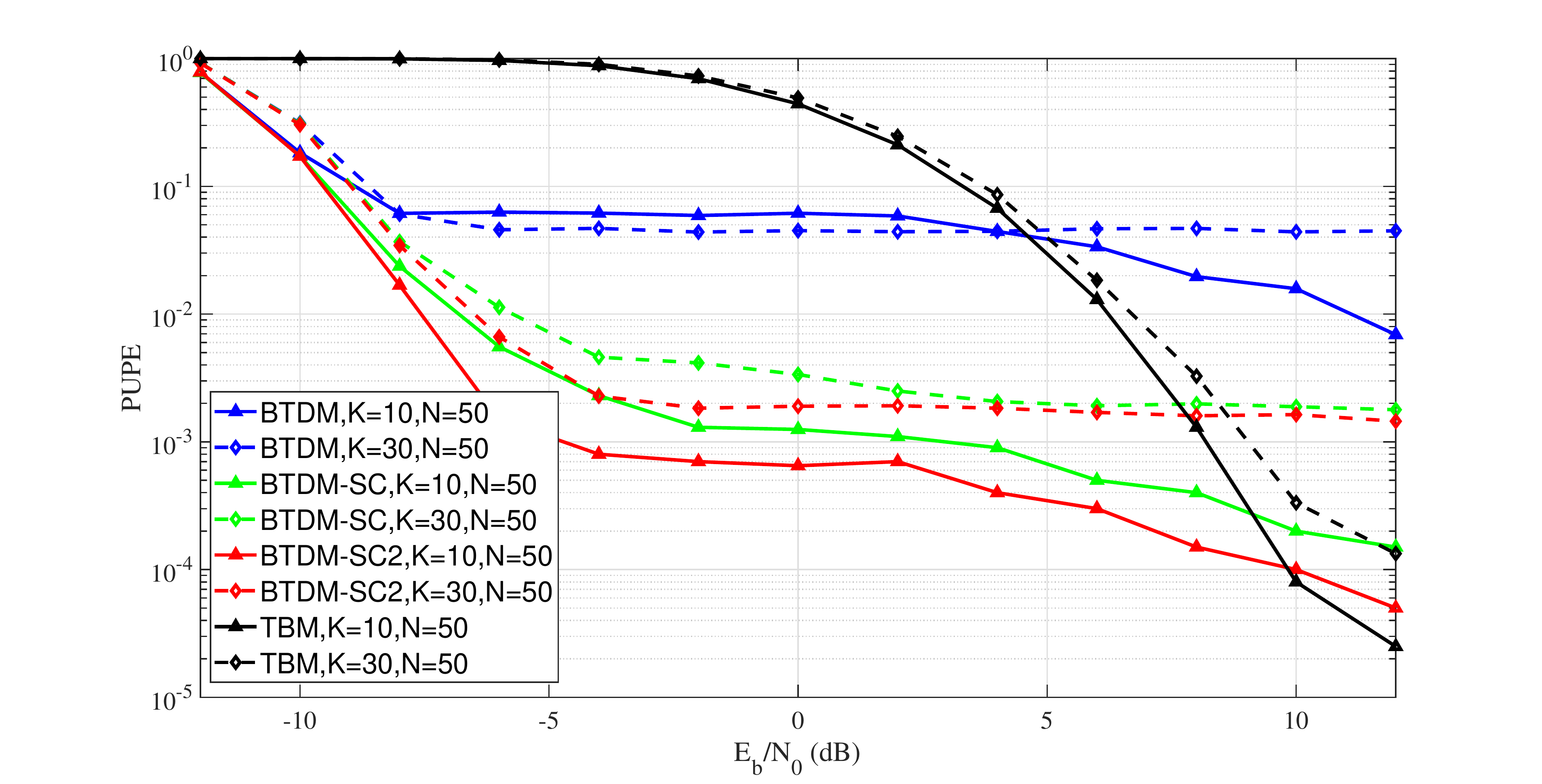}}
    \caption{The PUPE  of TBM and BTDM with $0\sim 2$ SC iterations. All simulations are implemented with the same transmission rate and  power.}\label{simulation1}
\end{figure}

Fig. \ref{simulation1} shows the PUPE performance of BTDM and TBM. From Fig. \ref{simulation1}, we can see  that the BTDM significantly outperforms TBM when $E_b/N_0\!<\!5dB$. \textcolor{black}{To further enhance the performance of BTDM, we employ successive cancellation (SC) strategy. Specifically, the BCH code is used for error detection and refining the incorrect messages by subtracting the correctly decoded messages from $\textit{\textbf{Y}}$ and re-decomposing the remaining tensor with BTD. } The green and red lines in Fig. \ref{simulation1} show the PUPE performance of BTDM with one and two SC iterations, respectively. Evidently,   SC  improves the performance even with  one iteration.
\begin{remark}
\textcolor{black}{
    For each dotted BTDM line in Fig. \ref{simulation1}, the number of active users is beyond the upper bound, thus  unrecoverable block terms always exist in (\ref{receivedsignal}), which contributes to the emergence of the error floor. In contrast, all solid BTDM lines with fewer users enjoy further decrease in PUPE  as $E_b/N_0$ increases.  To avoid error floors and accommodate more users in high $E_b/N_0$, we introduce a parallel technique for BTDM in next subsection.}
\end{remark}

\subsection{User Grouping and Parallel Demodulation}\label{subsectionparallel}

\begin{figure}[ht]
    \centerline{\includegraphics[width=9cm,height=4.8cm]{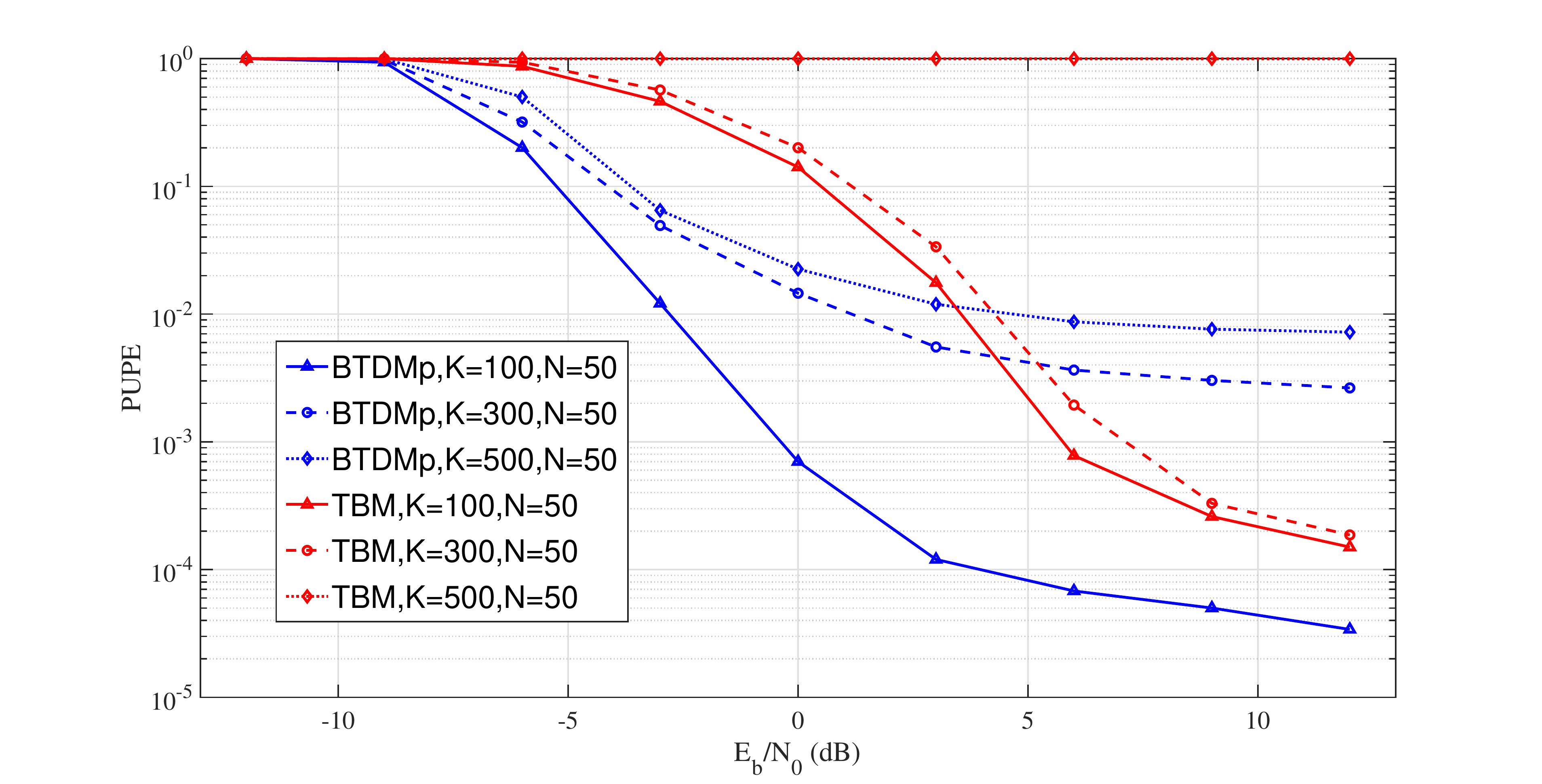}}
    \caption{The PUPE  of parallel BTDM and TBM.}\label{simulation2}
\end{figure}

To enable massive connectivity, users are divided into groups  associated with different channel resource blocks. Hence, the received signals can be demodulated in parallel. In Fig. \ref{simulation2}, the total number of channel uses is $T_c=7200$ and all channel resources are uniformly allocated among $G=10$ groups of users. The number of information bits and coded bits are given as $B_0\!=\!395$ and $B_{\text{BCH}}\!=\!440$. In each group, the tensor scale and channel uses are the same as in  Fig. \ref{simulation1}. For  TBM, the total number of channel uses is $T_c$ and the tensor size is $(T_1,T_2)\!=\!(90,80)$. It is shown in Fig. \ref{simulation2} that  BTDM can accommodate hundreds of random-access users with the user-grouping strategy. Compared to TBM, the parallel BTDM can accommodate more than 500 users, which is beneficial for scenarios with high payload and low $E_b/N_0$.


\subsection{Complexity}

The complexity of GNDL is $O(K^3((T_1+T_2)L+N)^3)$ for each iteration, which is dominated by solving a large linear system with singular value decomposition \cite{gndl}. The decoding procedures from factor matrices to information bits consist of three main steps, i.e., finding the dominant rows, solving equation (\ref{equation}) and demapping the entries of matrices to information bits. Hence, the complexity of the decoding process is $O(2K((T_1+T_2)L+(T_1+T_2)log(T)))$, which is much lower than  GNDL. Thus,  the demodulation features complexity $O(it_{\text{GNDL}}K^3((T_1+T_2)L+N)^3),$ where $it_{\text{GNDL}}$ is the number of iterations in GNDL. Similarly, the complexity of TBM with GNDL is $O(it_{\text{GNDL}}K^3(T_1+T_2+N)^3)$ \cite{btdpart2}. \textcolor{black}{Therefore, TBM, as a special case of BTDM with $L=1$, has lower complexity than BTDM.}

With the user-grouping strategy, the complexity of BTDM reduces to $O(it_{\text{GNDL}}K^3((T_1+T_2)L+N)^3/G^2)$. Therefore, the strategy significantly reduces the time complexity in decoding. The resulting complexity is inversely proportional to $G^2$, i.e., the square of the number of groups. Thus, the user-grouping BTDM has a substantially lower complexity than TBM when the number of groups becomes large.

\section{Conclusions}\label{sectionconclusion}

In this letter, we propose a new modulation scheme based on block term decomposition of tensors. A high-dimensional Grassmann constellation is designed  to match  the block terms and to improve the decoding performance through its rotation- and scale-invariant properties. Successive cancellation is used to further enhance the decoding performance when the number of active users exceeds the threshold required for unique decomposition. We apply  successive  cancellation  and  parallel demodulation  to cope with dense connectivity and decoding complexity. Numerical results validate the performance enhancements by our proposed tensor-BTD-based modulation.

\bibliographystyle{IEEEtran}
\bibliography{lib}

%
\IEEEpeerreviewmaketitle



\ifCLASSOPTIONcaptionsoff
  \newpage
\fi

\end{document}